\documentclass[twocolumn,showpacs,superscriptaddress,longbibliography,prb]{revtex4-1}

\usepackage{graphicx}
\usepackage{dcolumn}
\usepackage{amsmath}
\usepackage{color}
\usepackage{braket}
\usepackage{float}
\usepackage[american]{babel}

\newcommand{\SrCoVO}{SrCo$_2$V$_2$O$_8$}

\begin{document}

\title{From confined spinons to emergent fermions: Observation of elementary magnetic excitations in a transverse-field Ising chain}

\author{Zhe~Wang}
\email{e-mail: zhe.wang@physik.uni-augsburg.de}
\affiliation{Experimental Physics V, Center for Electronic
Correlations and Magnetism, Institute of Physics, University of Augsburg, 86135 Augsburg, Germany}

\author{Jianda~Wu}
\email{e-mail: jdwu@physics.ucsd.edu}
\affiliation{Department of Physics, University of California, San Diego, California 92093, USA}
\author{Shenglong~Xu}
\author{Wang~Yang}
\author{Congjun~Wu}
\affiliation{Department of Physics, University of California, San Diego, California 92093, USA}

\author{Anup~Kumar~Bera}
\altaffiliation[Present address: ]
{Solid State Physics Division, Bhabha Atomic Research Centre, Mumbai 400085, India}
\affiliation{Helmholtz-Zentrum Berlin f\"{u}r Materialien und
Energie, 14109 Berlin, Germany}

\author{A.~T.~M.~Nazmul~Islam}
\affiliation{Helmholtz-Zentrum Berlin f\"{u}r Materialien und
Energie, 14109 Berlin, Germany}

\author{Bella~Lake}
\affiliation{Helmholtz-Zentrum Berlin f\"{u}r Materialien und
Energie, 14109 Berlin, Germany}
\affiliation{Institut f\"{u}r Festk\"{o}rperphysik, Technische
Universit\"{a}t Berlin, 10623 Berlin, Germany}

\author{Dmytro~Kamenskyi}
\author{Papori~Gogoi}
\author{Hans~Engelkamp}
\affiliation{High Field Magnet Laboratory (HFML-EMFL), Radboud University, 6525 ED Nijmegen, The Netherlands}

\author{Nanlin Wang}
\affiliation{International Center for Quantum Materials, School of Physics, Peking University, 100871 Beijing, China}
\affiliation{Collaborative Innovation Center of Quantum Matter, Beijing, China}

\author{Joachim~Deisenhofer}
\affiliation{Experimental Physics V, Center for Electronic
Correlations and Magnetism, Institute of Physics, University of Augsburg, 86135 Augsburg, Germany}

\author{Alois~Loidl}
\affiliation{Experimental Physics V, Center for Electronic
Correlations and Magnetism, Institute of Physics, University of Augsburg, 86135 Augsburg, Germany}

\date{\today}

\begin{abstract}
We report on spectroscopy study of elementary magnetic excitations in an Ising-like antiferromagnetic chain compound \SrCoVO~as a function of temperature and applied transverse magnetic field up to 25 T. An optical as well as an acoustic branch of confined spinons, the elementary excitations at zero field, are identified in the antiferromagnetic phase below the N\'{e}el temperature of 5 K and described by a one-dimensional Schr\"{o}dinger equation. The confinement can be suppressed by an applied transverse field and a quantum disordered phase is induced at 7 T. In this disordered paramagnetic phase, we observe three emergent fermionic excitations with different transverse-field dependencies. The nature of these modes is clarified by studying spin dynamic structure factor of a 1D transverse-field Heisenberg-Ising (XXZ) model using the method of infinite time evolving block decimation. Our work reveals emergent quantum phenomena and provides a concrete system for testifying theoretical predications of one-dimension quantum spin models.
\end{abstract}

\maketitle

\section{Introduction}

Emergent states of matter in quantum magnets are characterized by their elementary excitations that can be induced and tuned in an external magnetic field.
Due to enhanced quantum fluctuations and reduced dimensionality, the low-dimensional quantum magnets host very often exotic quantum states of matter \cite{Giamarchi04, Zapf14, Punk14, Han12, Lake13, Mourigal13, Dalla-Piazza15, Wang16, Coldea10, Lake10, Goff95,LSWu16,Toskovic16}.
Especially for the one-dimensional (1D) spin systems, realization of novel quantum spin states is not only of lively experimental interest \cite{Lake13,Lake10,Mourigal13,Coldea10,LSWu16,Toskovic16} but also a constant focus of theoretical study \cite{Giamarchi04,Zapf14,Pereira06, Kohno09,Wu14,Bruognolo16},
since quantitative description of the elementary excitations can be provided by rigorous theoretical approaches and precise comparisons  between theoretical predictions and experimental results can be made \cite{Lake13,Lake10,Mourigal13,Coldea10,Kohno09,LSWu16,Toskovic16}.

\begin{figure}[t]
\centering
\includegraphics[width=90mm,clip]{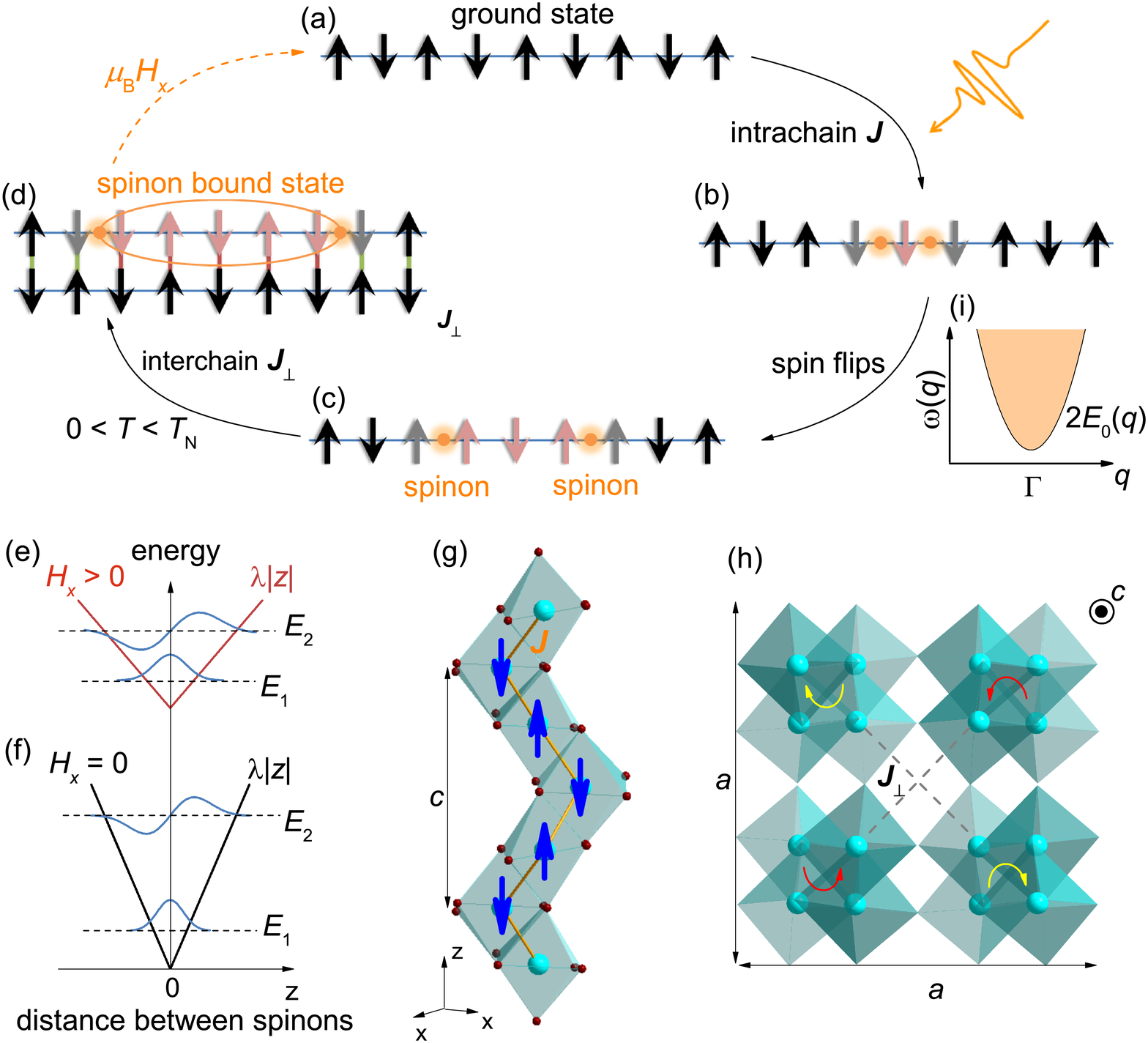}
\vspace{2mm} \caption[]{\label{Fig:Confinement_Scheme}
(a) Ground state of an antiferromagnetic Ising chain with antiparallel alignment of neighboring spins.
(b) A single spin-flip excitation composed of two spinons (orange dots).
(c) Spinons can propagate along the chain by subsequent spin flips due to $xy$ components of the exchange interactions.
(d) Two spinons form a spinon bound state due to the frustration of inter-chain exchange $J_\perp$.
(e) and (f) Potential energy increases linearly with distance between spinons $V(z)=\lambda |z|$ with $\lambda=2J_\perp\langle S_z\rangle^2/c$ and $c$ is the lattice constant along the chain direction, covering four spin sites in SrCo$_2$V$_2$O$_8$ [see (g) and (h)]. The two lowest-lying spinon bound states with energies $E_1$ and $E_2$ are schematically shown for zero and finite transverse fields $H_x$ with $\lambda(H_x=0)>\lambda(H_x>0)$. (g) Screw chain of edge-shared CoO$_6$ octahedra propagating along the crystallographic \emph{c} axis in SrCo$_2$V$_2$O$_8$ with antiferromagnetic order of Co$^{2+}$ spins along the \emph{c} direction stabilized below the N\'{e}el temperature $T<T_N$. (h) Viewing of left- and right-handed screw chains along the \emph{c} direction. The dominant interchain exchange $J_\perp$  is between the nearest-neighbor Co ions in the \emph{ab} plane, which are from the chains with the same chirality. $J_\perp > 0$ is antiferromagnetic.  (i) Parabolic dispersion relation of the energy threshold $2E_0$ in the reciprocal space close to the $\Gamma$ point ($q=0$) \cite{Shiba80a,Shiba80}.
}
\end{figure}

In the 1D spin-1/2 systems, the elementary magnetic excitations are spinons with fractional spin quantum number $S=1/2$ \cite{Faddeev81}.
Confinement of the fractional spinon excitations can be even realized \cite{Coldea10, Lake10, Morris14, Kimura07, Grenier15, Wang15a} as an analogy with quark confinement in particle physics \cite{Muta87}.
The concept of spinon confinement is illustrated in [Fig.~\ref{Fig:Confinement_Scheme}(a)-(d)]. In a classical picture, the ground state of a spin-1/2 Ising antiferromagnetic chain corresponds to an antiparallel alignment of neighboring spins [Fig.~\ref{Fig:Confinement_Scheme}(a)]. A single spin flip frustrates the intra-chain exchange interactions $J$ and creates two spinons, each with spin-1/2 [Fig.~\ref{Fig:Confinement_Scheme}(b)]. Subsequent spin flips can lead to propagation of the spinons along the chain [Fig.~\ref{Fig:Confinement_Scheme}(c)]. In the presence of weak inter-chain exchange interactions $J_\perp \ll J$ [Fig.~\ref{Fig:Confinement_Scheme}(d)], a 3D N\'{e}el antiferromagnetic order can be stabilized at low temperature $T<T_N$ and the separation of spinons from each other is unfavorable due to the frustrated inter-chain couplings. Thus the two spinons feel a confining potential, increasing with the distance between them, and form quantized spinon bound states [Fig.~\ref{Fig:Confinement_Scheme}(e)(f)] \cite{Shiba80a,Shiba80}.
Since the spinons are confined by the inter-chain couplings only in the antiferromagnetic phase, the magnetic phase transition can be thought as accompanied by a spinon confinement-deconfinement transition.

SrCo$_2$V$_2$O$_8$ is a representative realization of the paradigmatic 1D Heisenberg-Ising (XXZ) model. It crystallizes in the tetragonal structure with space group I4$_1$cd. Screw spin chains in SrCo$_2$V$_2$O$_8$ are constituted by edge-sharing CoO$_6$ octahedra and propagate along the crystallographic \emph{c} axis [Fig.~\ref{Fig:Confinement_Scheme}(g)] \cite{Bera14,He2006}. Each screw period consists of four Co$^{2+}$ ions with effective spin $S=1/2$, corresponding to the lattice constant \emph{c}. The unit cell contains four screw chains with either left- or right-hand chirality [Fig.~\ref{Fig:Confinement_Scheme}(h)]. Due to spin-orbit coupling and crystal field splitting, the Co$^{2+}$ spins have an Ising-like anisotropy along the \emph{c} axis \cite{Bera14,He2006}. The dominant exchange interaction $J$ is antiferromagnetic and between the nearest-neighbor Co$^{2+}$ spins in the chain. The inter-chain coupling $J_\perp$ is much smaller and dominated by exchange between the nearest-neighbor Co$^{2+}$ ions in the \emph{ab} plane that are from the neighboring chains with the same chirality [Fig.~\ref{Fig:Confinement_Scheme}(h)]. At $T_N=5$~K a long-range N\'{e}el-type antiferromagnetic order along the chain is stabilized [Fig.~\ref{Fig:Confinement_Scheme}(g)] \cite{Bera14,He2006,Canevet2013,Kawasaki2011}.

Below $T_N$ confinement of spinons is characterized by a series of quantized spinon bound states [Fig.~\ref{Fig:Confinement_Scheme}(f)] \cite{Wang15a}, which with spin $S=+1$ or $-1$ can be split in longitudinal magnetic field and exhibit linear field dependence \cite{Kimura07,Wang15a}. By applying a transverse magnetic field, the N\'{e}el temperature can be reduced according to the magnetic susceptibility measurements \cite{Bera14,He2006,Niesen14,Niesen13,Kimura2013}. At 2 K, a phase transition from the N\'{e}el antiferromagnetic (AFM) phase to a disordered paramagnetic (PM) phase is induced by a transverse field of 7 T \cite{Bera14,He2006}.

The reduction of N\'{e}el temperature suggests that the applied transverse magnetic field competes with the weak inter-chain couplings, and thus reduces the confining effects. This motivates us to utilize a transverse field to tune the spinon confinement. By continuously changing the transverse field, the confinement is expected to be suppressed with deconfined spinons observed.

Here we perform terahertz spectroscopy on the low-energy spin excitations in the 1D Ising-like antiferromagnetic-chain SrCo$_2$V$_2$O$_8$ as a function of transverse magnetic field, supplemented by studying spin dynamics of the corresponding 1D transverse-field XXZ model using the rigorous method of infinite time evolving block decimation (iTEBD). In zero field and below $T_N$, confinement of spinons of optical branch as well as of acoustic branch is observed and identified. Both the optical and the acoustic series of confined spinons can be described by a 1D Schr\"{o}dinger equation with linear confining potential. In small transverse fields, where the system remains antiferromagnetically ordered, the confining effects are found to be reduced: The confining potential becomes shallower with increasing field and the higher energy spinon bound states tend to collapse on the lowest level. At the transverse-field-induced order-disorder transition, we observe collapse of the quantized confined-spinon levels. Above this transition, the low-energy spin excitations in SrCo$_2$V$_2$O$_8$ are not deconfined spinons, but characterized by emergent fermions of the 1D transverse-field Ising class. We successfully identify the observed emergent fermions and their field dependencies by performing theoretical study of the 1D spin-1/2 XXZ model using the iTEBD method.

\section{Experimental details and theoretical method}

High-quality single crystals of SrCo$_2$V$_2$O$_8$ were grown using the floating-zone method and charactized by X-ray diffraction, neutron diffraction, and magnetization measurements \cite{Bera14}. Single crystals for the optical study were oriented using Laue diffraction and cut perpendicular to the tetragonal \emph{a} axis with a typical surface of $4\times4$~mm$^2$ and a thickness of 1 mm. High-field optical measurements were performed in the High Field Magnet Laboratory in Nijmegen. A Michelson interferometer with mercury lamp was used to generate electromagnetic waves in the terahertz frequency range which were detected by a silicon bolometer cooled at 1.6~K. The sample transmission spectrum was measured in the spectral range of 1.2-10~meV (0.3-2.5~THz) at magnetic fields up to 25~T. Magnetic fields were applied perpendicular to the crystallographic \emph{ac} plane and parallel to the propagation direction of the electromagnetic wave (Faraday configuration).

Standard infinite time evolving block decimation (iTEBD) method \cite{Vidal2007,Orus2008} was used to study spin dynamics of the 1D XXZ antiferromagnetic model in a transverse field. This method has been applied to other 1D systems \cite{Vidal2007,White2008} and turned out to be efficient to study 1D spin dynamics. Similar to the method of density matrix renormalization group, the iTEBD method assumes that the ground state (GS) of the spin system can be expressed as a matrix product state \cite{Vidal2007,Orus2008}. The two-point correlation function
\begin{equation}
C(r,t)=\left\langle S_y(r,t)S_y(0,0)+S_z(r,t)S_z(0,0)\right\rangle_{GS}
\end{equation}
is calculated to compare with experimental results.
The spin components $S_y$ and $S_z$ are coupled to the THz magnetic field and the applied external magnetic field is set along the $x$ direction.
The spin dynamic structure factor (DSF) can be obtained via Fourier transformation
\begin{equation}\label{Eq:DSF_definition}
S^{yy+zz}(q,\omega)=\int\limits_{-\infty}^{\infty}dr \int\limits_{-\infty}^{\infty}dt C(r,t)e^{i (\omega t -q r)},
\end{equation}
corresponding to the magnetic-dipole excitations observed in the optical measurements.
In our calculation, the bond dimension is taken to be $40$ and the discrete time step in the real time evolution is $0.02/J$.
The maximal time difference $|t-t'|$ is set to $32/J$, and we apply the linear prediction method \cite{White2008} to further improve the resolution.

\section{Experimental results}

\begin{figure}[t]
\centering
\includegraphics[width=85mm,clip]{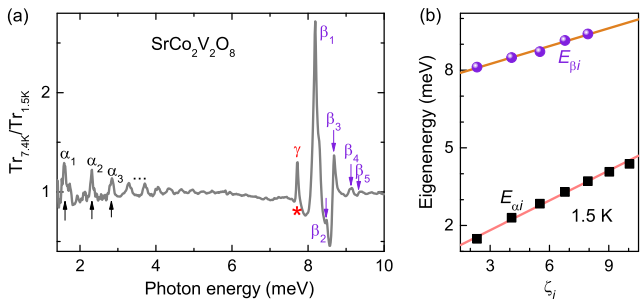}
\vspace{2mm} \caption[]{\label{Fig:Confined_LowField_HparaA}
(a) Ratio of transmission spectra that are measured at 7.4 K (above $T_N$) and at 1.5 K (below $T_N$). Two series of excitations $\alpha_i$ and $\beta_i$ are observed and indicated by the arrows, corresponding to confined spinons of acoustic and optical branch, respectively. Another mode $\gamma$ at 7.72 meV is marked by the star.
(b) Eigenenergies $E_{\alpha i}$ and $E_{\beta i}$ of the modes $\alpha_i$ and $\beta_i$, respectively, as a function of $\zeta_i$ the negative zeros of the Airy function \cite{McCoy78}. Linear fits according to the 1D Schr\"{o}dinger equation are shown by the solid lines [see Eqs.~(\ref{Eq:SpinnonSchroedingerEq}) and (\ref{Eq:SpinnonSchroedingerSolution})].
}
\end{figure}

\subsection{Confinement of acoustic and optical spinons in zero field}

A spin-flip excitation can be triggered by absorbing a photon, when the photon has the same energy as the two spinons created [Fig.~\ref{Fig:Confinement_Scheme}(a)(b)]. The resonance absorption of photons is manifested by a transmission minimum (or an absorption maximum) at the corresponding photon energy. After the creation, the two spinons can propagate along the chain [Fig.~\ref{Fig:Confinement_Scheme}(c)]. In presence of an effective linear confining potential $\lambda|z|$, the two-spinon bound state $\varphi(z)$ as a function of the distance $z$ between them can be described by the one-dimensional Schr\"{o}dinger equation
\begin{equation}\label{Eq:SpinnonSchroedingerEq}
    -\frac{\hbar^2}{\mu}\frac{d^2\varphi}{dz^2}+\lambda |z| \varphi=(E-2E_0)\varphi
\end{equation}
with $\hbar$ the Planck constant, $\mu$ an effective mass, and $2E_0$ the energy threshold for creating two spinons.
$\lambda=2J_\perp\langle S_z\rangle^2/c$ is an effective description of the inter-chain exchange interaction $J_\perp$, with the lattice constant \emph{c}. The solution of Eq.~(\ref{Eq:SpinnonSchroedingerEq}) is given by the linear dependence of $E_i$ on $\zeta_i$
\begin{equation}\label{Eq:SpinnonSchroedingerSolution}
    E_i=2E_0+\zeta_i\lambda^{2/3}\left(\frac{\hbar^2}{\mu}\right)^{1/3}
\end{equation}
where $i=1,2,3,...$ and $\zeta_i$ are given by the negative zeros of the Airy function $Ai(-\zeta_i)=0$ \cite{McCoy78}.

Figure~\ref{Fig:Confined_LowField_HparaA}(a) shows the ratio of two transmission spectra of SrCo$_2$V$_2$O$_8$ that are measured above and below the N\'{e}el temperature $T_N=5$~K. Spin excitations due to the magnetic phase transition are manifested by the peaks. One can see three groups of excitations that are denoted by $\alpha_i$, $\beta_i$, and $\gamma$. Eigenenergies $E_{\alpha i}$ of the $\alpha_i$ series follow the linear dependence on $\zeta_i$ as given in Eq.~(\ref{Eq:SpinnonSchroedingerSolution}) with $2E_0=0.68$~meV and  $\lambda^{2/3}\left(\frac{\hbar^2}{\mu}\right)^{1/3}=0.38$~meV as shown in Fig.~\ref{Fig:Confined_LowField_HparaA}(b).
This indicates that the $\alpha_i$ series is confined spinon excitations, which is consistent with previous study \cite{Wang15a}. In the same representation [Fig.~\ref{Fig:Confined_LowField_HparaA}(b)], one can see that the $\beta_i$ modes with higher eigenenergies $E_{\beta i}$ also follow a linear dependence and the linear fit is given by $2E_0=7.55$~meV and $\lambda^{2/3}\left(\frac{\hbar^2}{\mu}\right)^{1/3}=0.23$ meV.
Thus, the $\beta_i$ modes can be identified as another branch of confined spinons.
Compared to the $\alpha_i$ series, the $\beta_i$ modes have higher energy threshold and larger effective mass, if the same confining potential is assumed for both branches.

It will be shown in the following that we are able to assign the $\alpha_i$ modes as confined spinons of acoustic branch and the $\beta_i$ modes as of optical branch, in analogy with the concept of optical and acoustic phonons. In contrast, the mode $\gamma$ at 7.72~meV does not exhibit a series of excitations with higher energies and thus not fit into the same scheme as the $\alpha_i$ and $\beta_i$ modes.
It is worth noting that the $\beta_i$ modes are observed in the same energy range as the lower-lying optical phonon bands, making the higher energy levels difficult to resolve.

\subsection{Tuning the spinon confinement in low transverse magnetic field}

\begin{figure}[t]
\centering
\includegraphics[width=75mm,clip]{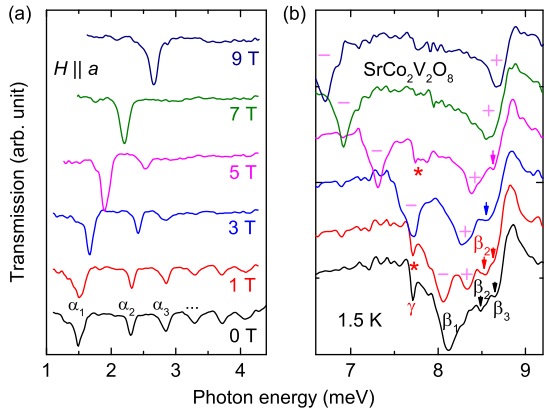}
\vspace{2mm} \caption[]{\label{Fig:Deconfined_LowField_HparaA}
Transmission spectra of (a) the acoustic $\alpha_i$ and (b) the optical $\beta_i$ spinons in applied transverse magnetic fields along the crystallographic \emph{a} axis ($H \parallel a$) up to 9 T. Series of absorption modes $\alpha_i$ can be observed at each field in (a). In (b) the zero-field mode $\beta_1$ splits into two modes in magnetic fields due to Zeeman interaction, as marked by the plus and minus signs. Below 7 T, shifting of the mode $\beta_2$ to higher energy with increasing fields is observed and indicated by the arrows. The $\gamma$ mode at 7.72 meV, marked by the star, has almost field-independent eigenenergy below 7 T and is not resolvable at 7 T or above.
}
\end{figure}

\begin{figure}[b]
\centering
\includegraphics[width=85mm,clip]{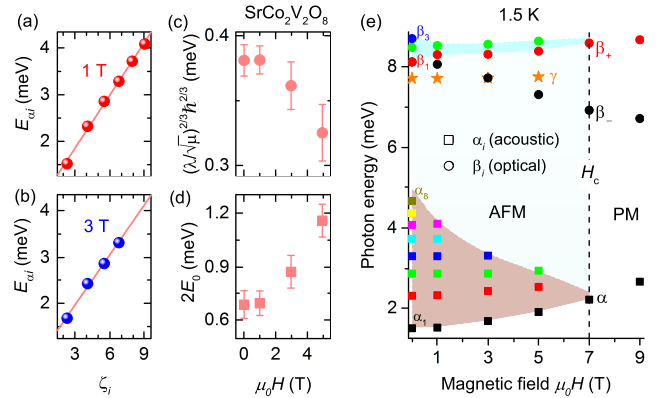}
\vspace{2mm} \caption[]{\label{Fig:Deconfined_LowField_FittingParameter}
(a) and (b) Eigenenergies $E_{\alpha i}$ of the absorption modes $\alpha_i$ as a function of $\zeta_i$ for 1 T and for 3 T. The lines are linear fits according to the 1D Schr\"{o}dinger equation with linear confining potential
[Eqs.~(\ref{Eq:SpinnonSchroedingerEq}) and (\ref{Eq:SpinnonSchroedingerSolution})]. (c) and (d) The fitting parameters $\lambda^{2/3}\left(\frac{\hbar^2}{\mu}\right)^{1/3}$  and $2E_0$ as functions of the transverse magnetic field.
(e) Eigenenergies of the acoustic $\alpha_i$, the optical $\beta_i$ spinons and the $\gamma$ mode as functions of the magnetic field. At 7 T, the higher levels collapse to the lowest-level mode, $\alpha$ for the acoustic and $\beta_+$ for the optical spinons; the $\gamma$ mode is suppressed; and only three modes $\alpha$, $\beta_+$, and $\beta_-$ remain. This indicates a field-induced phase transition from the antiferromagnetically (AFM) ordered to a paramagnetic (PM) quantum-disordered phase \cite{Bera14,He2006} at $H_c=7$~T.
}
\end{figure}

Figures \ref{Fig:Deconfined_LowField_HparaA}(a) and \ref{Fig:Deconfined_LowField_HparaA}(b) show transmission spectra of the acoustic $\alpha_i$ and optical $\beta_i$ spinons of SrCo$_2$V$_2$O$_8$ as a function of applied transverse magnetic field along the crystallographic \emph{a} axis ($H \parallel a$) up to 9~T. At zero field the series of transmission minima, corresponding to confined acoustic and optical spinons, can be identified in agreement with the temperature-dependent measurements [Fig.~\ref{Fig:Confined_LowField_HparaA}(a)].
For the acoustic branch [Fig.~\ref{Fig:Deconfined_LowField_HparaA}(a)], a series of transmission minima for 1 T is observed at almost the same energies as for 0 T, while at 3 T the resolved modes are shifted to higher energies [see also Fig.~\ref{Fig:Deconfined_LowField_FittingParameter}(e)]. Interestingly, the dependence of the resonance energies $E_i$ on $\zeta_i$ still follows the characteristic dependence at finite transverse fields, as shown in Figs.~\ref{Fig:Deconfined_LowField_FittingParameter}(a) and \ref{Fig:Deconfined_LowField_FittingParameter}(b). This indicates that the spinon confinement is still a valid description of the spin dynamics in low transverse fields. The linear fit determines the parameters $\lambda^{2/3}\left(\frac{\hbar^2}{\mu}\right)^{1/3}=0.36$~meV and $2E_0=0.87$~meV for 3 T [Fig.~\ref{Fig:Deconfined_LowField_FittingParameter}(b)], where the former decreased while the latter significantly increased compared to the results at zero field. Thus, by applying a transverse magnetic field, one not only changes the energy threshold of the confined spinons, but also tunes the profile of the confining potential.

As shown in Figs.~\ref{Fig:Deconfined_LowField_FittingParameter}(c) and \ref{Fig:Deconfined_LowField_FittingParameter}(d), the confining potential and threshold energy show a non-linear decrease and increase, respectively, with increasing transverse field. This is different from the situation of small longitudinal fields along the Ising axis, where the threshold energy exhibits Zeeman-type linear dependence on field and the confining-potential profile is almost unchanged \cite{Kimura07,Wang15a}, since the $U(1)$ symmetry of the spin Hamiltonian is broken in the transverse field.
The results in Figs.~\ref{Fig:Deconfined_LowField_FittingParameter}(c) and \ref{Fig:Deconfined_LowField_FittingParameter}(d) suggest that the confining potential becomes shallower in the transverse field.
In a shallower confining potential, as illustrated in Figs.~\ref{Fig:Confinement_Scheme}(e)(f), the difference between two neighboring
spinon bound states reduces. When the difference becomes very small, the higher level cannot be distinguished experimentally.
With increasing transverse field, the highest resolvable mode shifts to lower energy levels and, finally, when the second level ($\alpha_2$) collapses onto the first one ($\alpha_1$), the spinon hierarchy breaks down and the spinon confinement is completely suppressed.

\begin{figure*}[t]
\centering
\includegraphics[width=140mm,clip]{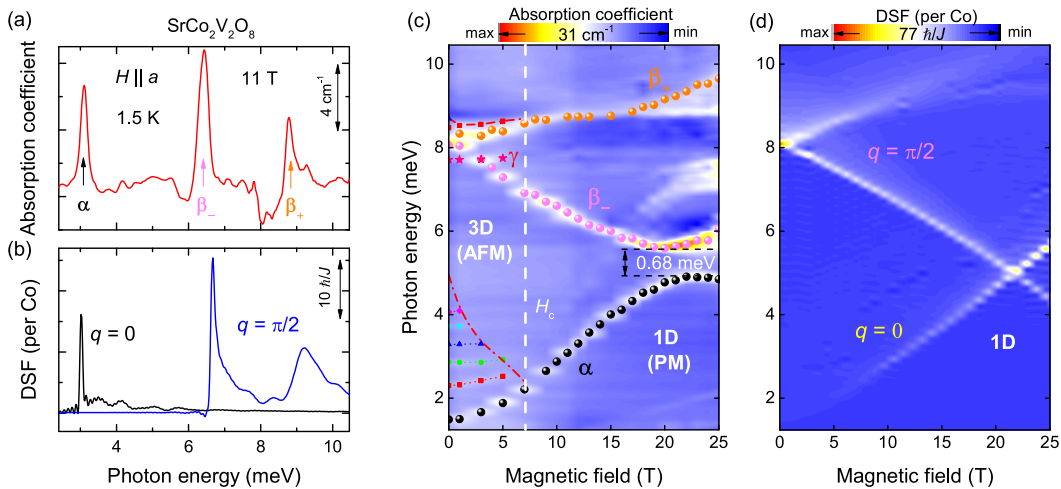}
\vspace{2mm} \caption[]{\label{Fig:EmergentFermion_HighField}
(a) Absorption-coefficient spectrum measured in transverse magnetic field of 11~T at 1.5~K, exhibiting three spin excitation modes $\alpha$, $\beta_-$, and $\beta_+$. The peak positions of the three modes are shown in (c) as a function of magnetic field (dots). (b) Spin dynamic structure factor (DSF) at $q=0$ and $q=\pi/2$ as a function of energy in the corresponding transverse magnetic field, simulated by the iTEBD method \cite{Vidal2007,Orus2008} based on the 1D spin-1/2 XXZ model in Eq.~(\ref{Eq:Ising-Heisenberg}). (c) Absorption coefficient and (d) DSF at $q=0$ and $q=\pi/2$ as a function of photon energy and transverse magnetic field. The vertical dashed line at $H_c=7$~T indicates the field-induced phase transition from the 3D antiferromagnetic (AFM) ordered to the 1D paramagnetic (PM) disordered phase \cite{Bera14,He2006}. In (c), the confined spinon levels $E_i$ shown in Fig.~\ref{Fig:Deconfined_LowField_FittingParameter}(e) are also marked (symbols and dotted lines), and a hybridization gap of 0.68 meV is indicated. The calculated results in (b) and (d) are given with parameters $J=3.55$ meV, $\Delta=2.00$, and $g_x=2.79$ (see text).
}
\end{figure*}

As shown in Fig.~\ref{Fig:Deconfined_LowField_HparaA}(b), the zero-field optical mode $\beta_1$ splits into two in the transverse fields due to Zeeman interaction, one shifting up while the other shifting down, as marked by the plus and minus signs. The higher-levels of confined optical spinons should be also split in the transverse field, since they belong to the same series. Due to strong phonon absorption at the same spectral range, fewer up-shifting confined modes of the optical branch can be resolved by the field-dependent measurements [marked by the arrows in Fig.~\ref{Fig:Deconfined_LowField_HparaA}(b)], while higher levels of the down-shifting modes are covered by the strong $\beta_1$ absorptions. Nevertheless, one can clearly observe some similar tuning behavior of the confined optical spinons as of the acoustic modes: The higher-level modes move towards the first level with increasing transverse field, although they all shift to higher energies with increasing fields; Above 3 T the $\beta_3$ mode cannot be distinguished from the $\beta_2$ mode, and at 7 T only the main modes ($\beta_+$ and $\beta_-$) can be observed indicating the suppression of the spinon confinement.

In addition, the $\gamma$ mode observed at 7.72 meV in zero field [Fig.~\ref{Fig:Confined_LowField_HparaA}(a)] shifts only slightly to higher energy and becomes broader with the increasing transverse fields. At 7~T, the $\gamma$ mode is already suppressed.Although the origin of this mode remains unclear, it is natural to relate the $\gamma$ mode to the 3D magnetic order because it is observed only below the magnetic phase transition.

The eigenenergies of the confined acoustic and optical spinons, and of the $\gamma$ mode are summarized in Fig.~\ref{Fig:Deconfined_LowField_FittingParameter}(e) as a function of the applied transverse magnetic field. Transverse-field tuning of the spinon confinement is obvious: Starting from zero field, both the acoustic and optical spinons are confined due to the inter-chain couplings and exhibit clear hierarchy of the spinon bound states. With increasing field, the inter-chain couplings are effectively reduced and the higher-level bound states are shifted downwards until the second-level collapses onto the first-level at 7 T. The breakdown of the hierarchy, suggesting the suppression of the inter-chain couplings, is an evident signature of the field-induced phase transition \cite{Bera14,He2006}.

\subsection{Emergent fermions of 1D transverse-field Ising class}

The above section documents the deconfining behavior of the confined spinons in a transverse magnetic field below the phase transition at $H_c=7$~T. Above $H_c$ in the quantum disordered phase, one would expect emergent elementary excitations that characterize spin state of the disordered phase. The characteristic excitations can be revealed by performing magneto-optic experiment in high magnetic fields.

A typical absorption-coefficient spectrum is shown in Fig.~\ref{Fig:EmergentFermion_HighField}(a) for a transverse field of 11 T above the field-induced phase transition. One can observe three absorption maxima at 3.1, 6.4, and 8.8 meV, corresponding to the $\alpha$, $\beta_-$, and $\beta_+$ modes, respectively. Their evolution with the transverse field is displayed in Fig.~\ref{Fig:EmergentFermion_HighField}(c) by a contour-plot of the absorption coefficient as a function of photon energy and transverse field, together with colored symbols highlighting their eigenenergies. The three modes $\alpha$, $\beta_-$, and $\beta_+$ are the dominant excitations that can be unambiguously followed in higher fields.

The $\alpha$ mode, evolving from the lowest-lying confined spinon excitation $\alpha_1$ [Figs.~\ref{Fig:Deconfined_LowField_HparaA}(a) and \ref{Fig:Deconfined_LowField_FittingParameter}(e)], increases in energy with increasing fields. The $\beta_-$ and $\beta_+$ modes correspond to the degenerate $\beta_1$ mode at zero field which is split in finite fields. At 22 T the $\alpha$ and $\beta_-$ modes do not cross but form a small hybridization gap of 0.68~meV [Fig.~\ref{Fig:EmergentFermion_HighField}(c)]. Above 22~T, the mode $\alpha$ shifts slightly downwards while the $\beta_-$ mode shifts upwards.

\begin{figure*}[t]
\centering
\includegraphics[width=170mm,clip]{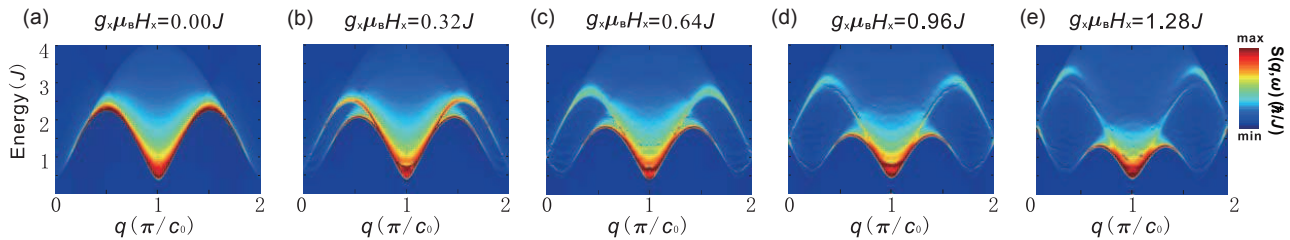}
\vspace{2mm} \caption[]{\label{Fig:S2_DynamicSF}
(a)-(e) Spin dynamic structure factor $S(q,\omega)$ [Eq.~(\ref{Eq:DSF_definition})] in the reciprocal space (0, $2\pi/c_0$) calculated up to $4J$ for the 1D Heisenberg-Ising XXZ model Eq.~(\ref{Eq:Ising-Heisenberg}) in various transverse magnetic fields $g_x \mu_B H_x$ of (a) 0, (b) 0.32, (c) 0.64, (d) 0.96, and (e) 1.28$J$ by the iTEBD method \cite{Vidal2007,Orus2008}. $c_0$ is the natural lattice constant of the 1D XXZ model and $J$ is the nearest-neighbor exchange interaction. The dynamic structure factor represents the dispersion relation of (a) spinons and (b)-(e) emergent fermions.
}
\end{figure*}

\section{Discussion}

In order to understand origin of the observed modes in high fields and clarify the two branches of confined spinons,
we use the standard iTEBD method \cite{Vidal2007,Orus2008} to study the spin dynamics of the spin-1/2 1D Heisenberg-Ising XXZ model, which describes the magnetic interactions of the cobalt spin chain. In a transverse magnetic field, the spin Hamiltonian is given by
\begin{equation}\label{Eq:Ising-Heisenberg}
    J \sum_{\left\langle i,j  \right\rangle}( S_{x,i}S_{x,j}+S_{y,i}S_{y,j}+\Delta S_{z,i}S_{z,j})- g_{x}\mu_B H_x \sum_{i}S_{x,i}
\end{equation}
where $\left\langle i,j  \right\rangle$ denotes the nearest neighbors, $J$ is the nearest-neighbor intra-chain antiferromagnetic exchange interaction, and $\Delta$ refers to an Ising-like anisotropy \cite{Bonner64,Lines63}. The last term is the Zeeman interaction term with the transverse magnetic field $H_x$ along the \emph{a} axis, the Land\'{e} \emph{g}-factor, and the Bohr magneton $\mu_B$.

The obtained spin dynamic structural factor (DSF) according to Eq.~(\ref{Eq:DSF_definition}) is shown for the whole Brillouin zone at various transverse magnetic fields in Fig.~\ref{Fig:S2_DynamicSF}. At zero field the known dispersion relation of spinons \cite{Bougourzi98,Lake10} is well reproduced in our calculations [Fig.~\ref{Fig:S2_DynamicSF}(a)]. The dispersion relation is symmetric with respect to $q=\pi$. Two minima of the lower-energy boundary are degenerate and located at $q=0$ and $q=\pi$, respectively. In a finite transverse field, the energy at $q=0$ starts to increase and the corresponding minimum shifts to larger $q$ values, while at $q=\pi$ the energy minimum is only slightly lowered. Moreover, a clear split of the dispersion relation occurs between $q=0$ and $q=\pi$ and becomes more significant with the increasing transverse fields. These features of the dispersion relation characterize emergent fermions of the transverse-field Ising-like chain \cite{Dmitriev02,Caux03}.

Terahertz spectroscopy is expected to probe the dispersion relation close to the $\Gamma$ point ($q=0$) in reciprocal space, since the wavelength of terahertz electromagnetic waves is much larger than the lattice constants of SrCo$_2$V$_2$O$_8$. To compare with the experimental results, in Fig.~\ref{Fig:EmergentFermion_HighField}(d) the obtained DSF at $q=0$ is shown as a function of energy and transverse magnetic field, which resembles the field dependence of the experimentally observed $\alpha$ mode. This indicates that the $\alpha$ fermionic excitations feature the spin dynamics at the $\Gamma$ point and the corresponding confined spinons observed at zero field belong to an acoustic branch.

To reconcile the experimental results, the DSF at $q=\pi/2$ is also shown in Fig.~\ref{Fig:EmergentFermion_HighField}(d) as a function of transverse field. An excitation mode close to 8 meV at zero field splits into two branches in the finite transverse fields, one increasing in energy while the other decreasing. Figure~\ref{Fig:EmergentFermion_HighField}(b) shows the DSF spectra of $q=0$ and $q=\pi/2$ which agrees very well with the absorption-coefficient spectrum of 11~T in Fig.~\ref{Fig:EmergentFermion_HighField}(a). The good agreement between experiment and theory is achieved with the parameters $J=3.55$~meV, $g_x=2.79$ and $\Delta=2.00$.
The $g_x$ factor smaller than the longitudinal $g_\parallel=5.5$ and the $\Delta>1$ value reveals an Ising-like anisotropy.\cite{Wang15a}
The exchange interaction is quite larger than the value of $J=2.8$~meV in BaCo$_2$V$_2$O$_8$,\cite{Kimura07} while the anisotropy of the two systems is very close to each other.

These results indicate that the experimentally observed $\beta_-$ and $\beta_+$ modes correspond to the emergent-fermion excitations at $q=\pi/2$ of the 1D transverse-field XXZ model, which are the low-energy spin excitations characteristic for the quantum disordered phase.
It is also clarified that the $\beta_i$ modes observed at zero field [Fig.~\ref{Fig:Confinement_Scheme}(a)] can be assigned as confined spinons of optical branch, in analogy with the concept of optical phonons.
At zero field the DSF at $q=\pi/2$ is much larger than that at $q=0$. This explains the stronger absorption of the $\beta_1$ mode compared to the $\alpha_1$ mode, as observed in Fig.~\ref{Fig:Confined_LowField_HparaA}(a).

The observation of the $q=\pi/2$ mode by optical spectroscopy indicates that the Brillouin zone is folded by a factor of four and translational invariance of the 1D XXZ model is broken due to the four spin-sites per unit cell in SrCo$_2$V$_2$O$_8$ [Fig.~\ref{Fig:Confinement_Scheme}(g)]. The zone-folding effect also explains the experimental observation of the hybridization gap at 22 T [Figs.~\ref{Fig:EmergentFermion_HighField}(c)]. In the extended Brillouin zone the crossing of the acoustic $\alpha$ and optical $\beta_-$ modes [Figs.~\ref{Fig:EmergentFermion_HighField}(d)] is protected by the translational symmetry, since they are located at different $q$ vectors. Due to the broken symmetry in the material, the transverse field can lead to non-zero off-diagonal elements of the interaction matrix, and thus mixes $\alpha$ and $\beta_-$ states and opens the hybridization gap.

With further increasing transverse fields, our calculation suggests that the lower-boundary at $q=\pi$ will be first gapless at 55~T, where a transverse-field Ising quantum critical point \cite{Dmitriev02,Caux03} is expected to emerge. At the quantum critical point, predicted dynamical features \cite{Zamolodchikov89,Wu14} are interesting to be testified in real materials.

\section{Summary}

Using high-resolution terahertz spectroscopy, we have studied low-energy spin dynamics of the Ising-like antiferromagnetic chain SrCo$_2$V$_2$O$_8$ in transverse fields up to 25~T. According to the terahertz electrodynamic response, the transverse field-tuned spin dynamics is clarified and it can be categorized into three regimes:

i) At zero field, the spin dynamics is characterized by confined spinons. We have identified confined spinons of an optical branch in addition to an acoustic branch. Both branches can be described by a 1D Schr\"{o}dinger equation with linear confining potential. Compared to the acoustic one, the optical branch is found at higher energy with larger effective mass.

ii) We have shown that the confinement of the optical and of the acoustic spinons can be tuned down by a small transverse field: The confining potential becomes shallower and the higher confinement levels are shifted closer to the lowest one. These properties can be viewed as deconfining behavior of the confined spinons, since the spinon series can be still described by the 1D Schr\"{o}dinger equation and the small transverse field can be viewed as a perturbation.

iii) We have also provided experimental evidence of the suppression of the spinon confinement by large transverse field, which is manifested by the collapse of the spinon hierarchy. This, however, does not result in deconfined spinons, but leads to the emergent-fermion excitations of the 1D transverse-field Ising class, because the transverse field is too strong to be treated as perturbation and even a phase transition from the antiferromagnetic to paramagnetic phase is induced. The evolution of emergent fermions in high transverse fields has been revealed by the terahertz spectroscopy and by studying low-energy spin dynamics of the 1D Ising-like XXZ model using the method of infinite time evolving block decimation.
A quantum critical point of transverse-field Ising class is predicted to occur at 55~T yet to be testified.

\begin{acknowledgments}
We would like to thank I. Affleck, S.-W. Cheong, B. Grenier, Zhangzhen He, T. Lorenz, A. Rosch, and S. Zvyagin for stimulating discussions.
We acknowledge partial support by the Deutsche Forschungsgemeinschaft via the Transregional Research Collaboration TRR 80: From Electronic Correlations to Functionality (Augsburg - Munich - Stuttgart) and the Project DE 1762/2-1, and by the Chinesisch-Deutsches Zentrum f\"{u}r Wissenschaftsf\"{o}rderung. The high field experiments were supported by the HFML-RU/FOM, member of the European Magnetic Field Laboratory (EMFL). J.W., S.X., W.Y., and C.W. acknowledge the support by NSF DMR-1410375 and AFOSR FA9550-14-1-0168. J.W. acknowledges the hospitality of Rice Center for Quantum Materials. C.W. acknowledges the supports from the President's Research Catalyst Awards CA-15-327861 from the University of California Office of the President, the National Natural Science Foundation of China (11328403), the CAS/SAFEA International Partnership Program for Creative Research Teams of China.
\end{acknowledgments}




\end{document}